\documentclass[pra,aps,groupaddress,prd]{revtex4}
\usepackage{amsfonts}
\usepackage{epsfig,amsmath,graphicx,amssymb,overpic,bm}
\usepackage{color,tikz,xcolor,extarrows}
\usetikzlibrary{arrows}

\def\be{\begin{equation}}
\def\ee{\end{equation}}
\def\bee{\begin{eqnarray}}
\def\ene{\end{eqnarray}}
\def\bes{\begin{subequations}}
\def\ees{\end{subequations}}

\newcommand{\bI}{{\bf I}}

\newcommand{\bR}{{\rm R}}
\newcommand{\mG}{\mathcal{G}}

\newcommand{\bQ}{{\bf Q}}
\newcommand{\bPT}{{\mathcal P}{\mathcal T}}
\newcommand{\bP}{{\mathcal P}}
\newcommand{\bT}{{\mathcal T}}
\newcommand{\bM}{{\rm M}}

\begin{document}

\title{Nonlocal general vector nonlinear Schr\"odinger equations: \\ Integrability, $\bPT$ symmetribility, and solutions}
\author{Zhenya Yan}
\email{zyyan@mmrc.iss.ac.cn}
\affiliation{Key Laboratory of Mathematics Mechanization, Institute of Systems
Science, AMSS, Chinese Academy of Sciences, Beijing 100190, China }
\date{\today}

\begin{abstract}
{\bf Abstract}\,\, A family of new one-parameter $(\epsilon_x=\pm 1)$ nonlinear wave models (called $\mG_{\epsilon_x}^{(nm)}$ model) is presented, including both the local $(\epsilon_x=1)$ and new integrable nonlocal $(\epsilon_x=-1)$ general vector nonlinear Schr\"odinger (VNLS) equations with the self-phase, cross-phase, and multi-wave mixing modulations. The nonlocal $\mG_{-1}^{(nm)}$ model is shown to possess the Lax pair and infinite number of conservation laws for $m=1$. We also establish a connection between the $\mG_{\epsilon_x}^{(nm)}$ model and some known models. Some symmetric reductions and exact solutions (e.g., bright, dark, and mixed bright-dark solitons) of the  representative nonlocal systems are also found. Moreover, we find that the new general two-parameter $(\epsilon_x, \epsilon_t)$ model (called $\mG_{\epsilon_x, \epsilon_t}^{(nm)}$ model) including the $\mG_{\epsilon_x}^{(nm)}$ model is invariant under the $\bPT$-symmetric transformation and the $\bPT$ symmetribility of its self-induced potentials
 is discussed for the distinct two parameters $(\epsilon_x, \epsilon_t)=(\pm 1, \pm 1)$.

\vspace{0.1in}
\noindent {\it Keywords:}  Nonlocal general VNLS equations; Lax pair; conservation laws; $\bPT$ symmetribility; symmetry reductions; exact solutions; solitons.

\end{abstract}

%\pacs{05.45.Yv, 02.30.Ik, 42.65.Tg}
\maketitle

\baselineskip=12pt

%%%%%%%%%%%%%%%%%%%%%%%%%%%%%%%%%%%%%%%%%%%%%%%%%%%%%%%%%%%%%%%%%

\vspace{-0.2in}

\section{Introduction}

 The nonlinear Schr\"odinger (NLS) equation is a typical physical model appearing in many fields of nonlinear science, including nonlinear optics, Bose-Einstein condensates, plasma physics, biology, deep ocean, finance, etc.~\cite{ab,nls1,nls2,nls3,bec,oc,yan2010}. The cubic NLS equation is completely integrable and admits multi-soliton solutions, breather solutions, periodic-wave solutions, and rogue waves~\cite{ab, vnls2,ma, ak, ps}. Many nonlinear wave equations, in particular, soliton equations, are local in mathematical physics (see Ref.~\cite{ab} and reference therein). Recently, a new nonlocal NLS equation $iq_t(x,t)=q_{xx}(x,t)\pm 2q^2(x,t)q^{*}(-x,t)$ was shown to be integrable~\cite{am}. Particularly, its self-induced potential, $V(x,t)=\pm 2q(x,t)q^{*}(-x,t)$, was shown to be $\bPT$-symmetric for the fixed time. Here the parity reflection operator $\mathcal{P}:$\, $x\rightarrow -x$ is linear whereas the time reflection operator $\mathcal{T}:$\, $t\to -t,\, i\rightarrow -i$ is anti-linear~\cite{pt1}. Nowadays, the $\bPT$ symmetry plays a more and more role in many fields of science (see, e.g., Refs.~\cite{pt1,pt2,pt3} and references therein). More recently, we~\cite{yanaml15} presented a new two-parameter local and nonlocal vector NLS equations ($\bQ_{\epsilon_x, \epsilon_t}^{(n)}$ model), in which the $\bQ_{-1, 1}^{(n)}$ model was shown to be integrable nonlocal vector NLS equations with the $\bPT$-symmetric self-induced potentials for the fixed time. It is still of important significance to seek for integrable nonlinear wave models in the soliton theory and integrable systems~\cite{ab,vnlsb}. Up to now, there are a few nonlinear systems which are both integrable and $\bPT$-symmetric. In this letter we will introduce and investigate in detail a new nonlocal general model, which is integrable and contains  the $\bPT$-symmetric self-induced potentials.

\section{A new nonlocal  general vector model: integrability and reductions}

{\it Nonlinear model.}---In this letter, we introduce a new general one-parameter $(\epsilon_x)$ model (shortly called $\mG_{\epsilon_x}^{(nm)}$ model)
\vspace{-0.03in} \bee  \label{vnlsp}
i\bQ_t(x,t)=-\bQ_{xx}(x,t)+2 \bQ(x,t)\bQ^{\dag}(\epsilon_xx, t)\bM\bQ(x,t),\quad  i=\sqrt{-1}
\ene
containing the local $(\epsilon_x=1)$ and {\it novel} nonlocal $(\epsilon_x=-1)$ general VNLS equations, where $x, t\in \mathbb{R}$,\, the subscripts denote partial derivatives with respect to corresponding variables, $\epsilon_{x}=\pm 1$ can be regarded as the  identity $(+)$ and $\bPT$ ($-$) symmetric parameters, $\bQ(x,t)=(q_{ij}(x,t))_{n\times m}$ is an $n\times m$ slowly varying complex-valued amplitude matrix, $\bQ^{\dag}(\epsilon_x x, t)$ denotes the transpose conjugate of $\bQ(\epsilon_x x, t)$,  $\bM=\bM^{\dag}=(M_{ij})_{n\times n}$ is an $n\times n$ constant Hermite matrix with $|M|\not=0$. We now consider the simple case $m=1$, the self-induced potentials $2\bQ^{\dag}(-x, t)\bM\bQ(x,t)$ in $\mathcal{G}_{-1}^{(n1)}$ model may {\it not} be real-valued, but self-induced potentials $2\bQ^{\dag}(x,t)\bM\bQ(x,t)$ is real-valued in $\mathcal{G}_{+1}^{(n1)}$ model. The quasi-power defined by $G_{\epsilon_x}(t)=\int_{-\infty}^{+\infty}\bQ^{\!\dag}(\epsilon_xx,t)\bQ(x,t)dx$ is conserved during evolution.
The total power of Eq.~(\ref{vnlsp}) is defined by $P_{\epsilon_x}(t)=\int_{-\infty}^{+\infty}\bQ^{\dag}(x,t)\bQ(x,t)dx$. It can be shown that the power $P_{-1}(t)$ is in general {\it not} conserved but the power $P_{+1}(t)$ is conserved during evolution since $dP_{\epsilon_x}(t)/(dt)=-2i\int_{-\infty}^{+\infty}\!\!dx|\bQ(x,t)|^2[\bQ^{\dag}(\epsilon_xx,t)\bM\bQ(x,t)-
\bQ^{\!\dag}(x,\!t)\bM\bQ(\epsilon_xx,t)]$.

{\it Lax pair.}---To seek for Lax pair of system (\ref{vnlsp}), we start from the linear iso-spectral problem (Lax pair)~\cite{vnlsb}
\vspace{-0.03in}\bes\label{lax} \bee
\label{lax1}\Psi_x=W_{n+m}(x, t,\lambda)\Psi,\quad W_{n+m}(x, t,\lambda)=-i\lambda \Sigma_3+U, \hspace{1.35in} \vspace{0.1in}\\
\label{lax2}\Psi_t=V_{n+m}(x, t,\lambda)\Psi,\quad V_{n+m}(x, t,\lambda)=-2i\lambda^2\Sigma_3+2\lambda U-i(U_x+U^2)\Sigma_3,\quad
\ene\ees
where $\Psi=\Psi(x,t)=(\psi_1(x,t),\psi_2(x,t),...,\psi_{n+m}(x,t))^{\rm T}$ is a complex column eigenvector, $\lambda\in \mathbb{C}$ is an iso-spectral parameter, the generalized Pauli matrix $\Sigma_3$  and potential matrix $U=U(x,t)$ are given by
\bee \nonumber
 \Sigma_3=\left(\begin{matrix} \bI_n  &  0_{n\times m} \vspace{0.05in}\\ 0_{m\times n} & -\bI_m \end{matrix} \right), \qquad
 U(x,t)= \left(\begin{matrix}  0_n &  \bQ(x,t) \vspace{0.05in}\\  \bR(x,t) & 0_m \end{matrix} \right), \,\,
\ene
where $I_n$ and  $ 0_n$ are $n\times n$ unity and zero matrixes, respectively,  $0_{n\times m}$ is a $n\times m$ zero matrix, and $\bR(x,t)=(r_{ij}(x,t))_{m\times n}$ is a $m\times n$ complex-valued  function matrix.  The compatibility condition of Lax pair (\ref{lax1})-(\ref{lax2}) $\Psi_{xt}=\Psi_{tx}$, that is, the zero curvature equation $W_{n+m,t}-V_{n+m,x}+[W_{n+m}, V_{n+m}]=0$,  yields a system of $2nm$-component nonlinear wave models~\cite{vnlsb}
\vspace{-0.05in}\bes \label{vnls}\bee\label{vnlsa}
i\bQ_t(x,t)=-\bQ_{xx}(x,t)+2 \bQ(x,t)\bR(x,t)\bQ(x,t), \quad\,\, \\
-i\bR_t(x,t)=-\bR_{xx}(x,t)+2 \bR(x,t)\bQ(x,t)\bR(x,t).\quad\,\,
\label{vnlsb}\ene\ees

In the following we consider system  (\ref{vnls}) for the special potential matrix $\bR(x,t)$. The symmetric reduction $\bR(x,t)=\bQ^{\dag}(x,t)\bM$ with the constant matrix $\bM=\bM^{\dag}$ and $|M|\not=0$~\cite{yang10} of system (\ref{vnls}) yields the general vector NLS equations
\vspace{-0.05in} \bee\label{nls0}
i\bQ_t(x,t)=-\bQ_{xx}(x,t)+2\bQ(x,t)\bQ^{\dag}(x,t)\bM\bQ(x,t), \quad
\ene
which corresponds to Eq.~(\ref{vnlsp}) with $\epsilon_x=1$. Eq.~(\ref{nls0}) can further yield the general coupled NLS equations ($m=1$) for the general Hermite matrix $\bM$~\cite{yang10}, as well as the Manakov system ($n=2,\, m=1$) and vector NLS equation~\cite{vnls} for the special Hermite matrix $\bM=\pm \bI_{n}$.

We nowadays choose a new one-parameter $(\epsilon_x)$ symmetric reduction
\bee\label{con}
 \bR(x,t)=\bQ^{\dag}(\epsilon_xx, t)\bM, \qquad (\bM=\bM^{\dag},\quad |M|\not=0, \quad  \epsilon_x=\pm 1)
\ene
such that system (\ref{vnls}) becomes the above-introduced new nonlocal system (\ref{vnlsp}). Therefore, we have shown that system (\ref{vnlsp}) admits the Lax pair (\ref{lax1}) and (\ref{lax2}) with $\bR(x,t)$ given by symmetric constraint (\ref{con}).

In what follows we investigate the $\mG_{\epsilon_x}^{(nm)}$ model (\ref{vnlsp}) with $\epsilon_x=\pm 1$ and $m=1$ in details, in which $\bQ(x,t)=(q_1(x,t), q_2(x,t),\cdots, q_n(x,t))^T$ is a complex-valued column vector.  Eq.~(\ref{vnlsp}) with $n=m=1$ (i.e., $\mG_{\epsilon_x}^{(11)}$ model) yields the usually local $(\epsilon_x=1)$~\cite{nls1} and nonlocal $(\epsilon_x=-1)$ NLS equation~\cite{am} in the unified form
\bee\label{nls1}
 iq_{1, t}(x,t)\!=\!-q_{1, xx}(x,t)\!+\!2M_{11}q_1^2(x,t)q_1^{*}(\epsilon_x x, t),\quad  M_{11}\in \mathbb{R},
\ene
which is completely integrable, where the star denotes the complex conjugate. Eq.~(\ref{vnlsp}) with $n=2,\, m=1$ ($\mG_{\epsilon_x}^{(21)}$ model) yields the local $(\epsilon_x=1)$~\cite{yang10} and {\it new} nonlocal $(\epsilon_x=-1)$ vector NLS equations
\bee\label{nls2} \begin{array}{rl}
iq_{j,t}(x,t)= & -q_{j,xx}(x,t)+2[M_{11} q_1(x,t)q_1^{*}(\epsilon_xx,t)+M_{22}q_2(x,t)q^{*}_2(\epsilon_x x,t) \vspace{0.1in}\\
 &\qquad\qquad\quad\quad\quad +M_{12}^{*}q_1(x,t)q_2^{*}(\epsilon_xx,t) +M_{12}q_2(x,t)q_1^{*}(\epsilon_xx, t)]q_j(x,t), \, (j=1,2) \qquad
  \end{array}\ene
where $M_{jj}\in \mathbb{R}\, (j=1,2)$,  $M_{jj}$ and $\lambda_{kk}\, (k\not=j)$ denote the self-phase modulation and cross-phase modulation effects for the component $q_{j}$, respectively,
$M_{12}$ and $M_{12}^{*}$ are the four wave mixing effects.

 For the special parameters $\{\epsilon_x,\, M_{ij}\}$,  $\mG_{\epsilon_x}^{(21)}$ model (\ref{nls2}) can reduce to some known models: i) we fix $\epsilon_x=1$. when $M_{11}=M_{22}=\sigma,\, M_{12}=0$,  system (\ref{nls2}) is just the well-known Manakov system~\cite{vnls}; when $M_{11}=-a=-M_{22},\, M_{12}=0$,  system (\ref{nls2}) becomes the mixed coupled NLS equations~\cite{vnlsm}; when $M_{11}=-a,\, M_{22}=-c,\, M_{12}=-b^{*}$,  system (\ref{nls2}) becomes the general coupled NLS equations~\cite{yang10}. (ii) we fix
 $\epsilon_x=-1$. When $M_{11}=M_{22}=\sigma,\, M_{12}=0$,  system (\ref{nls2}) is just the known nonlocal vector NLS equations reported recently in Ref.~\cite{yanaml15}; otherwise we obtain the {\it new nonlocal} vector NLS equations [cf. system (\ref{nls2}) with $\epsilon_x=-1$].

{\it Reductions.}---We now further reduce system (\ref{vnlsp}) by considering the property of Hermite matrix $\bM$.  Since $\bM$ is a Hermite (self-adjoint) matrix and $|M|\not=0$, then it has $n$ non-zero real eigenvalues $\lambda_j\, (j=1,2,...,n)$ and there exists an $n\times n$ unitary matrix ${\bf B}$ (i.e., ${\bf B}{\bf B}^{\dag}={\bf B}^{\dag}{\bf B}={\bf I}_n$) such that $\bM={\bf B}\Lambda {\bf B}^{\dag}$, where $\Lambda$ is a real diagonal matrix with the diagonal elements being $\lambda_j\, (j=1,2,3...,n)$.
We now make the transformation in system (\ref{vnlsp})
\bee
 \bQ(x,t)={\bf B}\Phi(x,t),
\ene
where $\Phi(x,t)=(\phi_{ij}(x,t))_{n\times m}$,  as a consequence, we have
\bee\label{sys-d}
 \bQ^{\dag}(\epsilon_xx, t)\bM\bQ(x,t)=\Phi^{\dag}(\epsilon_xx, t){\bf B}^{\dag}\bM{\bf B}\Phi(x,t)=\Phi^{\dag}(\epsilon_xx, t)\Lambda\Phi(x,t),
\ene
such that system (\ref{vnlsp}) reduces to
\vspace{-0.03in} \bee  \label{vnlsr}
i\Phi_t(x,t)=-\Phi_{xx}(x,t)+2 \Phi(x,t)\Phi^{\dag}(\epsilon_xx, t)\Lambda\Phi(x,t).
\ene

If we further make the scaling transformation
$\Phi(x,t)=\hat\Lambda\hat\Phi(x,t)$ with $\hat\Lambda={\rm diag}(1/\sqrt{|\lambda_1|},\, 1/\sqrt{|\lambda_2|}, ..., 1/\sqrt{|\lambda_n|})$, then system (\ref{vnlsr}) becomes
\bee
\label{vnlsc}
i\hat\Phi_t(x,t)=-\hat\Phi_{xx}(x,t)+2\hat\Phi(x,t)\Phi^{\dag}(\epsilon_xx, t)\hat\bI_n\hat\Phi(x,t),\qquad  \hat\bI_n={\rm diag}({\rm sgn}(\lambda_1),\, {\rm sgn}(\lambda_2), ..., {\rm sgn}(\lambda_n)),
\ene
Thus the transformation between systems (\ref{vnlsp}) and  (\ref{vnlsc}) can be written as
\bee\label{tran}
 \bQ(x,t)={\bf B}\hat\Lambda\hat\Phi(x,t).
 \ene
(i) If $\hat\bI_n=\pm \bI_n$ and $\epsilon_x=1$, then system (\ref{vnlsc}) becomes the know one~\cite{vnlsb}; otherwise, it is a mixed system;
(ii) If $\hat\bI_n=\pm \bI_n$,\, $m=1$, and $\epsilon_x=-1$, then system (\ref{vnlsc}) becomes the nonlocal focusing ($+)$ or defocusing $(-)$ vector NLS equations presented recently~\cite{am,yanaml15}; otherwise, it is a new mixed system.

{\it Conservation laws.}---In the following we apply the Lax pair (\ref{lax1}) and (\ref{lax2}) with constraint (\ref{con}) to seek for the infinite number of conservation laws of system (\ref{vnlsp}) with $m=1$.
We introduce the following $n$ new complex functions~\cite{wadati}
\bee
 \omega_j(x,t)=\frac{\psi_j(x,t)}{\psi_{n+1}(x,t)}\qquad (j=1,2,...,n) \ene
on the basis of $n+1$ eigenfunctions $\psi_j(x,t)$ of Lax pair (\ref{lax1})-(\ref{lax2}) with  condition (\ref{con}) and $m=1$ such that we find that $\omega_j(x,t)$ satisfy  $n$-component Riccati equations
from Eq.~(\ref{lax1})
\vspace{-0.03in}\bee
 \omega_{j,x}(x,t)\!=-\!\omega_j(x,t)\!\sum_{s,l=1}^nM_{kl}q_k^{*}(\epsilon_xx,t)\omega_l(x,t)
  -2i\lambda \omega_j(x,t)\!+\!q_j(x,t),\qquad\qquad
\label{ri}\ene
where we have used $r_j(x,t)=\sum_{k=1}^nM_{kj}q_k^{*}(\epsilon_xx, t)$.  To determine the functions $\omega_{j}(x,t)$ we assume that they are of the form
 \bee\label{omega}
\begin{array}{l} \omega_j(x,t)=\sum_{k=0}^{\infty}\frac{\omega_{j}^{(s)}(x,t)}{(2i\lambda)^{s+1}}.
\end{array}
\ene
We substitute Eq.~(\ref{omega}) into Eq.~(\ref{ri}) and compare the coefficients of same terms $(2i\lambda)^{s}\, (s=0,1,2,...)$ to determine these functions $\omega_j(x,t)$ in terms of potentials $q_j(\epsilon_x x,t)$ and $q_j(x,t)$
\bee \label{w1}\begin{array}{l}
 \omega_{j}^{(0)}(x,t)=q_j(x,t), \,\,\, \omega_{j}^{(1)}(x,t)=-q_{j,x}(x,t),\,\,\, \omega_{j}^{(2)}(x,t)=q_{j,xx}(x,t)-q_{j}(x,t)\sum_{k,l=1}^nM_{kl}q_{l}(x,t)q_k^{*}(\epsilon_xx, t),  \vspace{0.1in}\\
  \omega_{j}^{(s+1)}(x,t)\!=-\sum_{i,l=1}^nM_{li}q_l^{*}(\epsilon_xx, t)\!\!\sum_{k=1}^{s-1}\omega_{j}^{(k)}(x,t)\omega_{i}^{(s-k)}(x,t)
 -\omega_{j,x}^{(s)}(x,t), \,\, (s=2,3,...)
\end{array}\ene

It follows from Lax pair (\ref{lax1}) and (\ref{lax2}) with condition (\ref{con}) and $m=1$ that we have
\bee \nonumber \begin{array}{l}
 (\ln |\psi_{n+1}|)_x\!=\!i\lambda+F(x,t), \qquad
 (\ln |\psi_{n+1}|)_t\!=\!2i\lambda^2+ G(x,t),
 \end{array}\ene
where  $F(x,t)=\sum_{k,j=1}^nq_k^{*}(\epsilon_xx,t)M_{kj}\omega_j(x,t)$ and $G(x,t)=\sum_{k,j=1}^nM_{kj}\{[2\lambda q_k^{*}(\epsilon_xx, t)-i\epsilon_x q_{k,x}^{*}(\epsilon_xx, t)]\omega_j(x,t)+iq_j(x,t)q_k^{*}(\epsilon_xx, t)\}$ with $q_{k,x}^{*}(\epsilon_xx, t)=\partial q_{k}^{*}(\xi, t)/(\partial \xi)|_{\xi=\epsilon_xx}$.
Thus, the compatibility condition, $(\ln |\psi_{n+1}|)_{xt}=(\ln |\psi_{n+1}|)_{tx}$ leads to the conservation laws
\bee\label{conver} \partial_t F(x,t)= \partial_x G(x,t).
\ene

We substitute expression (\ref{omega}) into Eq.~(\ref{conver}) and compare the coefficients of same terms $(2i\lambda)^{-j}\, (j=1, 2,3,...)$ yields the infinite number of conservation laws [cf. Eq.~(\ref{w1})]. For example, we have the first two local $(\epsilon_x=1)$ and nonlocal $(\epsilon_x=-1)$ conservation laws of system (\ref{vnlsp}) with $m=1$
\bee \nonumber
\partial_t\sum_{k,j=1}^nM_{kj}q_j(x,t)q_k^{*}(\epsilon_xx, t)=i\partial_x\sum_{k,j=1}^nM_{kj}[q_{j,x}(x,t)q_{k}^{*}(\epsilon_xx, t)
 -\epsilon_x q_{j}(x,t)q_{k,x}^{*}(\epsilon_xx, t)].
 \ene
\bee\nonumber
\partial_t\!\!\sum_{k,j=1}^n\!\!\! M_{kj}q_j(x,t)q_{k,x}^{*}(\epsilon_xx, t)=i\partial_x\!\!\sum_{k,j=1}^n\!\!\! M_{kj}[q_{j,x}(x,t)q_{k,x}^{*}(\epsilon_xx, t)
  \!\!-\!\!\epsilon_x q_{j}(x,t)q_{k,xx}^{*}(\epsilon_xx, t)\!\!-\!\!\epsilon_x q_{j}^2(x,t)q_{k}^{*2}(\epsilon_xx, t)].
 \ene
where $q_{k,x}^{*}(\epsilon_xx, t)=q_{k,\xi}^{*}(\xi, t)|_{\xi=\epsilon_xx}$.

\vspace{0.1in}
{\it The two-parameter $(\epsilon_x, \epsilon_t)$ extension of the model (\ref{vnlsp}).}---In fact, we can also extend the model (\ref{vnlsp}) to the more general form (simply called $\mG_{\epsilon_x, \epsilon_t}^{(nm)}$ model)
\vspace{-0.03in} \bee  \label{vnlsp-g}
i\bQ_t(x,t)=-\bQ_{xx}(x,t)+2 \bQ(x,t)\bQ^{\dag}(\epsilon_xx, \epsilon_tt)\bM\bQ(x,t),\quad  i=\sqrt{-1}
\ene
where $\epsilon_x=\pm 1,\, \epsilon_t=\pm 1$ and other matrix $\bM$ and $\bQ(x,t)$ are the same as ones in system (\ref{vnlsp}). When $\epsilon_t=1$, system (\ref{vnlsp-g}) reduces to the integrable model (\ref{vnlsp}). But when $\epsilon_t=-1$, system (\ref{vnlsp-g}) is also new and its integrable properties (e.g., Lax pair and conservation laws)
will be discussed in another literature.

\section{$\bPT$-symmetric invariance and self-induced potentials}

It is easy to see that system (\ref{vnlsp-g}) is invariant under the $\bPT$-symmetric transformation ($\bP: x\to -x$;\, $\bT: t\to -t,\, i\to -i$), that is, if $\bQ(x,t)$ is a solution of system (\ref{vnlsp-g}), then its $\bPT$ counterpart $\bPT\bQ(x,t)=\bQ^{*}(-x,-t)$ also solve the same system.

System (\ref{vnlsp-g}) with $m=1$ can be rewritten as $i\bQ_t(x,t)={\bf H}_n(\epsilon_xx, \epsilon_tt)\bQ(x,t)$, where the Hamiltonian matrix operator ${\bf H}_n$ with self-induced potential $2\bQ^{\dag}(\epsilon_xx,t)\bM\bQ(x,t)$~\cite{self} is given by
\bee \label{hn}
  {\bf H}_n(\epsilon_xx, \epsilon_tt)=[-\partial_x^2+2\bQ^{\dag}(\epsilon_xx,\epsilon_tt)\bM\bQ(x,t)]\bI_n
 \ene
In what follows we consider the $\bPT$ symmetribility of the matrix operator under the basic definition of $\bPT$ symmetry~\cite{pt1}.

 (i) When $\epsilon_x=-\epsilon_t=-1$, we have the self-induced potential $V_{n-+}(x,t)=2\bQ^{\dag}(-x,t)\bM\bQ(x,t)$ and find
\bee \nonumber 
V_{n-+}(x, t)\not\equiv V_{n-+}(-x, t), \,\,
V_{n-+}(x, t)=V_{n-+}^{*}(-x, t), \,\,
V_{n-+}(x, t)\not\equiv V_{n-+}^{*}(x, -t),\,\,
 V_{n-+}(x, t)\not\equiv V_{n-+}^{*}(-x, -t),
\ene
in terms of the Hermiticity of $\bM$, that is, the self-induced  potential $V_{n-+}(x,t)$ is not necessarily  $\bP$-, $\bT$-, and $\bPT$-symmetric. But it is $\bPT$-symmetric for any solution $\bQ(x,t)$ of system (\ref{vnlsp}) if $t$ is regarded as a fixed real parameter.

 (ii) When $\epsilon_x=\epsilon_t=1$, we know the self-induced potential $V_{n++}(x,t)=2\bQ^{\dag}(x,t)\bM\bQ(x,t)$ and find
\bee \nonumber 
V_{n++}(x, t)\not\equiv V_{n++}(-x, t), \,\,
V_{n++}(x, t)\not\equiv V_{n++}^{*}(-x, t), \,\,
V_{n++}(x, t)\not\equiv V_{n++}^{*}(x, -t),\,\,  V_{n++}(x, t)\not\equiv V_{n++}^{*}(-x, -t),
\ene
for $\bQ(x,t)$, that is,  the self-induced  potential $V_{n++}(x,t)$  is not necessarily $\bP$-, $\bT$-, and $\bPT$-symmetric. But if $\bQ(x,t)$ is an even column function matrix about space, then we have  $V_{np++}(x, t)=V_{np++}^{*}(-x, t)$, that is, it is $\bPT$-symmetric for the fixed parameter $t$.

(iii) When $\epsilon_x=-\epsilon_t=1$, we know the self-induced potential $V_{n+-}(x,t)=2\bQ^{\dag}(x,-t)\bM\bQ(x,t)$ and find
 \bee \nonumber
 V_{n+-}(x, t)\not\equiv V_{n-+}(-x, t), \,\,
V_{n+-}(x, t)\not\equiv V_{n-+}^{*}(-x, t), \,\,
V_{n+-}(x, t)= V_{n+-}^{*}(x, -t),\,\,  V_{n+-}(x, t)\not\equiv V_{n+-}^{*}(-x, -t),
\ene
that is,   the self-induced  potential $V_{n+-}(x,t)$ is $\bT$-symmetric and not necessarily $\bP$- and $\bPT$-symmetric.

(iv) When $\epsilon_x=\epsilon_t=-1$, we know the self-induced potential $V_{n--}(x,t)=2\bQ^{\dag}(-x,-t)\bM\bQ(x,t)$ and find
\bee \nonumber
V_{n--}(x, t)\not\equiv V_{n--}(-x, t), \,\,
V_{n--}(x, t)\not\equiv V_{n--}^{*}(-x, t), \,\,
V_{n--}(x, t)\not\equiv V_{n--}^{*}(x, -t),\,\,
 V_{n--}(x, t)=V_{n--}^{*}(-x, -t),
\ene
that is,  the self-induced potential $V_{n--}(x,t)$ is $\bPT$-symmetric and and not necessarily $\bP$- and $\bT$-symmetric.

\section{Multi-linear form and symmetry reductions}

 We now consider some symmetry reductions of system (\ref{vnlsp}) with $\epsilon_x=-1,\, m=1$. If $q_j(-x, t)=\epsilon_jq_j(x,t)$ with $\epsilon_j=\pm 1$, that is, they are even or odd functions for $x$, then nonlocal system (\ref{vnlsp}) reduces to the local system (\ref{nls0}) whose solutions have been studied (see, e.g.,  Refs.~\cite{yang10, vnls-s, vnls-s2} for $n=2$).

{\it Mutil-linear form.}---We make the general rational transformations~\cite{hirota} in Eq.~(\ref{vnlsp}) with $\epsilon_x=-1$ and $m=1$
\bee \label{bi1}
 q_j(x,t)=\frac{g_j(x,t)}{f(x,t)}, \quad f(x,t),\, g_j(x,t)\in\mathbb{C}[x,t]
\ene
where $f(x,t),\, g_j(x,t)\in\mathbb{C}[x,t]$ such that we have the system of (bi, tri)-linear equations
\bee
\begin{array}{l} {\rm Bilinear \,\, eq.:}\,\,\, (iD_t+D_x^2-\mu) g_j(x,t)\cdot f(x,t)=0,\vspace{0.1in} \\
{\rm Trilinear \,\, eq.:}\,\, f^{*}(-x, t)(D_x^2-\mu)f(x,t)\cdot f(x,t)
   =2f(x,t){\bf G}^{\dag}(-x, t)\bM{\bf G}(x,t), \end{array}
\label{bi2}\ene
where $\mu\in \mathbb{C}$, ${\bf G}(x,t)=(g_1(x,t), g_2(x,t),...,g_n(x,t))^T$ and ${\bf G}^{\dag}(-x, t)$ is the transpose conjugate of ${\bf G}(-x,t)$, $D_t$ and $D_x$ are both Hirota's bilinear operators defined by~\cite{hirota} $D_t^mD_x^n f\cdot g=(\partial_t-\partial_{t'})^m(\partial_x-\partial_{x'})^n[f(x,t)g(x,t)]|_{x=x',t=t'}$.
If ${\bf G}^{\dag}(-x, t)\bM{\bf G}(x,t)$ is a real-valuable function and $f(x,t)$ is an even function, then we can assume $f(x,t)\in\mathbb{R}[x,t]$ in from
Eq.~(\ref{bi1}), which leads to $f^{*}(-x, t)=f(x,t)$.Thus Eq.~(\ref{bi2}) becomes a bilinear equation.
 \bee
\begin{array}{l}  (iD_t+D_x^2-\mu) g_j(x,t)\cdot f(x,t)=0,\quad
 (D_x^2-\mu)f(x,t)\cdot f(x,t)=2{\bf G}^{\dag}(-x, t)\bM{\bf G}(x,t), \end{array}
\nonumber \ene

{\it Direct symmetry reductions.}---Now we apply the direct reduction method~\cite{am, yanaml15,ck} to study the symmetry reduction of system (\ref{vnlsp}) with $m=1$
\vspace{-0.03in}\bee \label{si1a}
\begin{array}{l}
q_j(x,t)=\frac{p_j(z)}{\sqrt{2t}}e^{i\mu\log|t|/2}, \,\,\, z(x,t)=x/\sqrt{2t},
\end{array}\ene
where $\mu\in\mathbb{R}$, $p_j(z)\in \mathbb{C}[z],\, (j=1,2,...,n)$, and $ x,t\in \mathbb{R}$. As a consequence, we find the local and nonlocal ordinary differential system  with variable coefficients for $p_j(z)$
\vspace{-0.03in} \bee \label{si1b}
  p_{j,zz}-(\mu+i)p_j(z)-izp_{j,z}(z)-2p_j(z)\sum_{k,s=1}^nM_{ks}p_s(z)p_k^{*}(\epsilon_x z)=0,
\ene
 where $\hat{z}=\epsilon_x x/\sqrt{2t}$. Notice that i) when $\epsilon_x=1$, we know $\hat{z}=z$ and $p_j^{*}(\hat{z})=p_j^{*}(z)$ such that the symmetry reduction becomes the usual local result; ii) when $\epsilon_x=-1$, we have $\hat{z}=-z$ and $p_j^{*}(\hat{z})=p_j^{*}(-z)$ such that the symmetry reduction given by Eq.~(\ref{si1a}) and (\ref{si1b})  becomes
 the nonlocal case.

Similarly, we consider the statioary solution of system (\ref{vnlsp}) with $m=1$
$q_j(x,t)=p_j(x)e^{i\mu t},\, (j=1,2,...,n)$
where $\mu\in \mathbb{R}$,\, $p_j(x)\in \mathbb{C}[x]$, and $ x,t\in \mathbb{R}$. Thus the functions $p_j(x)$ satisfy the ordinary differential system with constant coefficients
 \vspace{-0.05in}\bee\label{si2b}
   p_{j,xx}-\mu p_j(x)-2p_j(x)\sum_{k,s=1}^nM_{ks}p_s(x)p_k^{*}(\epsilon_x x)=0,
\ene
More symmetry redcutions of system~(\ref{vnlsp}) with any $m$ may be found by using Lie classical and non-classical symmetric methods~\cite{sm}.

 {\it Exact solutions.}---For the nonlocal system (\ref{nls2}) with $\epsilon_x=-1$, we will study it some solutions. We make the transformation~\cite{vnls-s2}
\bee \label{trans}
 q_1(x,t)=p_1(x,t)+M_{12}p_2(x,t),\quad q_2(x,t)=-M_{11}p_2(x,t)
 \ene
  such that system (\ref{nls2}) with $\epsilon_x=-1$ reduces to
\bee\label{nls-s} \begin{array}{rl}
ip_{j,t}(x,t)= & -p_{j,xx}(x,t)+2M_{11} [p_1(x,t)p_1^{*}(-x,t)+M_dp_2(x,t)p^{*}_2(-x,t)]p_j(x,t), \, (j=1,2) \qquad
  \end{array}\ene
where $M_d=M_{11}M_{22}-|M_{12}|^2\not=0$.  If $p_2(x,t)=C p_1(x,t)$ with $C$ being a constant, then we can reduce system (\ref{nls-s}) to one nonlocal NLS equation~\cite{am}
\bee\label{rnls}
 ip_{1,t}(x,t)= & -p_{1,xx}(x,t)+2M_{11}(1+M_d|C|^2)p_1^2(x,t)p^{*}_1(-x,t),
\ene
whose solutions were obtained in Ref.~\cite{yanaml15,am} such that we can obtain the solutions of system (\ref{nls2}) with $\epsilon_x=-1$ using the transformation (\ref{trans}).
Particularly, if $1+M_d|C|^2=0$ for some parameter $C$, then Eq.~(\ref{rnls}) becomes a complex linear heat equation $ip_{1,t}(x,t)=-p_{1,xx}(x,t)$, which admits the plane wave solutions $p_1(x,t)=\sum_{j=1}^{\infty}\exp[ik_j(x-k_jt)]$ for $k_j\in \mathbb{R}$.

Here we give other periodic wave solutions of nonlocal system (\ref{nls2}) with $\epsilon_x=-1$
\bes\label{s1}\bee
q_{1,sc}(x,t)=\sqrt{\frac{\beta^2(1-m^2)+\mu_1}{2M_{11}}}{\rm sn}(\beta x, k)e^{i\mu_1 t} +M_{12}\sqrt{-\frac{\beta^2(1+k^2)+\mu_1}{2M_{11}M_d}}{\rm cn}
(\beta x, k)e^{i(\beta^2k^2+\mu_1)t}, \\
q_{2,sc}(x,t)=-M_{11}\sqrt{-\frac{\beta^2(1+m^2)+\mu_1}{2M_{11}M_d}}{\rm cn}(\beta x, k)e^{i(\beta^2k^2+\mu_1)t},
\ene\ees
\bes\bee
q_{1,sd}(x,t)=\sqrt{\frac{\beta^2(k^2-1)+\mu_1}{2M_{11}}}{\rm sn}(\beta x, k)e^{i\mu_1 t} +M_{12}\sqrt{-\frac{\beta^2(1+k^2)+\mu_1}{2M_{11}M_d}}{\rm dn}(\beta x, k)e^{i(\beta^2+\mu_1)t}, \\
q_{2,sd}(x,t)=-M_{11}\sqrt{-\frac{\beta^2(1+k^2)+\mu_1}{2M_{11}M_d}}{\rm dn}(\beta x, k)e^{i(\beta^2+\mu_1)t},
\ene\ees
\bes \label{s3}\bee
q_{1,cd}(x,t)=\sqrt{\frac{\mu_1-\beta^2}{2M_{11}(1-k^2)}}{\rm cn}(\beta x, k)e^{i\mu_1 t} +M_{12}\sqrt{\frac{\beta^2(1-2k^2)+\mu_1}{2M_{11}M_d(k^2-1)}}{\rm dn}(\beta x, k)e^{i[\beta^2(1-k^2)+\mu_1]t}, \\
q_{2,cd}(x,t)=-M_{11}\sqrt{\frac{\beta^2(1-2k^2)+\mu_1}{2M_{11}M_d(k^2-1)}}{\rm dn}(\beta x, k)e^{i[\beta^2(1-k^2)+\mu_1]t},
\ene\ees
\bes\label{s4}\bee
 q_{1,scd}(x,t)=\frac{(1+M_{12}h)[2Ak \beta {\rm cn}(\beta x, k)+i\nu{\rm sn}(\beta x,k)]}
 {2kB \beta+2\sqrt{k^2B^2\beta^2+4A^2M_{11}(1+M_dh^2)}{\rm dn}(\beta x, k)}e^{i\beta^2(k^2-2)/2 t}, \\
 q_{2,scd}(x,t)=-\frac{M_{11}h[2Ak\beta{\rm cn}(\beta x,k)+i\nu{\rm sn}(\beta x,k)]}
 {2Bk \beta+2\sqrt{k^2B^2\beta^2+4A^2M_{11}(1+M_dh^2)}{\rm dn}(\beta x, k)}e^{i\beta^2 (k^2-2)/2 t},
\ene\ees
where $\mu_1,\, \beta,\, A,\, B,\, h$ are all real constants, $k\in (0,\, 1)$ is the elliptic modulus, and $\nu=k\beta\sqrt{4M_{11}^2(1-k^2)-B^2\beta^2k^4/[M_{11}(1+M_dh^2)]}$. In particularly, when $k\to 0$ and $k\to 1$, it follows from Eqs.~(\ref{s1})-(\ref{s4}) that we have the trigonometric function solutions and solitary wave solutions of nonlocal system (\ref{nls2}) with $\epsilon_x=-1$, respectively. The family of solutions  (\ref{s3}) and (\ref{s4}) are $\bPT$-symmetric.

The nonlocal system (\ref{nls2}) with $\epsilon_x=-1$ and system (\ref{vnlsp-g}) also possess other types of solutions such as breather solutions, rational solitons and rogue waves, which will be studied using the generalized Darboux transformation~\cite{gdt} and other approaches in another literature because of the page limit.

\section{Conclusions}

In conclusions, we have first introduced a two-parameter model including both the known local and new nonlocal general VNLS equation. The nonlocal general VNLS equations $\mG^{(nm)}_{-1, 1}$ are shown to admit the Lax pair and infinite number of conservation laws for $m=1$. The model $\mG^{(nm)}_{\epsilon_x, \epsilon_t}$ was shown to be of the $\bPT$-symmetric invariance.
The $\bPT$ symmetribility of the self-induced potentials in the model $\mG^{(n1)}_{\epsilon_x, \epsilon_t}$ is studied for the
distinct parameters $\epsilon_x$ and $\epsilon_t$. Particularly, based on some transformations we can reduce the nonlocal general VNLS equations to another nonlocal VNLS equations containing the known focusing and defocusing
nonlocal VNLS equations~\cite{yanaml15} and new mixed nonlocal VNLS equations. Their multi-linear forms and symmetry reductions are given. Moreover, we mainly find exact solutions of the nonlocal general two-component NLS equations containing double-periodic wave solutions, trigonometric function solutions and solitary wave solutions. The idea used in this letter may also be extended to find other new integrable nonlocal models.

\vspace{0.1in} \noindent {\bf Acknowledgments}

\vspace{0.1in}
The author thanks the referee for the valuable suggestions and comments. 
 This work was supported by the NSFC under Grant No. 11571346 and the Youth Innovation Promotion Association CAS.

%%%%%%%%%%%%%%%%%%%%%%%%%%%%%%%%%%%%%%%%%%%%%%%%%%%%%%%%%%%%%%

%\begin{thebibliography}{99}

%\vspace{-0.25in}

\end{document}